\documentclass[12pt]{article}
\usepackage{amssymb}
\usepackage{epsfig}
\input epsf
\begin{document}
\title{
\hfill{}\\
\hfill{}\vspace*{0.5cm}\\
\sc
Photonic crystals of coated metallic spheres
\vspace*{0.3cm}}
\author{ {\sc Alexander 
Moroz}\thanks{www.amolf.nl/research/photonic\_materials\_theory}
\thanks{
Address 
after Sept. 1, 1999:
I. Institut f\"{u}r Theoretische Physik,  Jungiusstrasse 9, 
Universit\"{a}t Hamburg, D-20355 Hamburg, Germany}
\vspace*{0.3cm}}
\date{
\protect\normalsize
\it FOM Institute AMOLF, 
Kruislaan 407,  1098 SJ Amsterdam,  The Netherlands}

\maketitle

\begin{center}
{\large\sc abstract}
\end{center}
It is shown that simple face-centered-cubic (fcc) structures of 
both metallic and coated metallic spheres are ideal candidates to achieve
a tunable complete photonic bandgap (CPBG) for optical wavelengths using
currently available experimental techniques. 
For coated microspheres with the coating width to plasma 
wavelength ratio $l_c/\lambda_p \leq  10\%$
and the coating and host refractive indices $n_c$ and $n_h$, respectively,  
between $1$ and $1.47$, one can always find a
sphere radius $r_s$ such that the relative gap width $g_w$ 
(gap width to the midgap frequency ratio) is larger
than $5\%$ and, in some cases, $g_w$ can exceed $9\%$.
Using different coatings and supporting liquids, the width and 
midgap frequency of a CPBG can be tuned considerably.

\vspace*{0.6cm}

\noindent PACS numbers:  42.70.Qs - Photonic bandgap materials \hfill\\
\noindent PACS numbers:  82.70.Dd - Colloids\hfill

\vspace*{1.9cm}

\thispagestyle{empty}
\baselineskip 20pt
\newpage


%

\noindent 
{\em Introduction}. -
Photonic crystals are characterized by a periodically modulated 
dielectric constant. Some of such structures 
occur in nature, for instance, opals and nanostructured colour 
wings of butterflies \cite{OG}.
There is a  common belief that in the near future photonic crystals 
will give us the same control over photons as
 ordinary crystals give us over electrons \cite{Y}. 
At the same time, photonic structures are of great 
promise to become a laboratory for testing fundamental processes involving 
interactions of radiation with matter in novel conditions. 
This promise originates from the fact that, in analogy  to the case of an 
electron moving in a periodic potential, certain photon frequency
modes 
within a photonic crystal can become forbidden, independent of the photon 
polarization and direction of propagation  - a complete photonic bandgap 
(CPBG) \cite{HCS,YGL}. Consequently, the density of states (DOS) and
the local DOS (LDOS) of photons are significantly changed compared to 
their vacuum value (see \cite{AMe} for the exact results in one-dimensional
photonic crystals). If the LDOS is sufficiently smooth, the 
spontaneous emission (SE) rate $\Gamma$ of atoms and molecules embedded 
in a photonic crystal is directly proportional to the LDOS \cite{Tip}.
On the other hand, if the LDOS exhibits sharp features (as a function
of frequency and position in the unit cell), one 
expects the Wigner-Weisskopf
approximation \cite{WW} to break down and novel phenomena to occur, 
such as non-Markovian behaviour and non-exponential SE 
accompanied by the  change of 
the spectrum from a single Lorentzian peak into a two-peaked structure
 \cite{By,LZM}.

Unfortunately, the problems in the fabrication of three-dimensional
CPBG structures increase rapidly
with decreasing wavelengths for which a CPBG is required -
mainly because of the simultaneous requirements on the modulation 
(the total number and the length of periodicity steps) and
dielectric contrast. In order to achieve a CPBG below infrared 
wavelengths, the modulation is supposed to be on the scale of optical 
wavelengths or even  smaller and, as for any CPBG structure, 
has to be achieved with roughly ten periodicity steps in each direction, 
a task  currently beyond the reach of reactive ion and chemical 
etching techniques. Fortunately, such a modulation occurs naturally
in colloidal crystals formed by monodisperse colloidal suspensions 
of microspheres. The latter are known to self-assemble into 
three-dimensional  crystals with excellent long-range order 
on the optical scale \cite{RSS},
removing the need for complex and costly microfabrication.
They form a face-centered-cubic (fcc)
or (for small sphere filling fraction) a body-centered-cubic (bcc) 
lattice \cite{RSS}. Thus, it comes as no surprise that the best
photonic crystals in the visible (although without any CPBG) 
are colloidal based fcc structures \cite{WV}.
The latter are purely dielectric structures composed of spheres with
the dielectric constant $\varepsilon_s$ embedded in a host 
with the dielectric constant $\varepsilon_h$.
For such structures the dielectric contrast $\delta$ is defined as 
$\delta=\mbox{max}\,(\varepsilon_h/\varepsilon_s,
\varepsilon_s/\varepsilon_h)$. 
Then $\delta \stackrel{\textstyle >}{\sim} 8.2$ 
is required to open a CPBG \cite{BSS,MS}.
Colloidal crystals suffer from the same kinds of defects as
ordinary electronic crystals. Therefore, practical crystals 
should have such a dielectric contrast  which for an ideal
crystal yields a  CPBG with the gap width-to-midgap frequency ratio 
(the relative gap width), $g_w=\triangle\omega/\omega_c$,
of at least $5\%$ - to leave a margin
for gap edge distortions due to impurities and yet to have a CPBG 
useful for applications. Then, for an fcc structure, 
$\delta\stackrel{\textstyle >}{\sim}12$ 
is required \cite{BSS,MS} which makes fabrication of photonic 
crystals with an operational CPBG at optical wavelength seemingly 
hypothetical \cite{Lev}.  
The requirements on $\delta$ are less restrictive (although still
rather strong in the visible)
for colloidal crystals with a diamond structure \cite{HCS}. The latter, 
however, have yet to be fabricated.
As a result of the above, no three-dimensional CPBG structure below the 
infrared wavelengths  \cite{Lev,Lin} have been fabricated thus far.

Recently we have shown \cite{AlM} that a way to avoid the requirements 
on $\delta$ is to use spheres with
a Drude-like behaviour of the dielectric function
\begin{equation}
\varepsilon_s(\omega) = 1 - \omega_p^2/\omega^2,
\label{drudeps}
\end{equation}
where $\omega_p$ is called the plasma frequency \cite{BH}.
In the following, $r_s$ is the sphere radius and 
$\lambda_p=2\pi c/\omega_p$ 
the plasma wavelength, $c$ being the speed of light in vacuum.
For notational simplicity we often refer to the spheres having such a 
dielectric function (\ref{drudeps}) as metallic spheres, although we 
are aware that (i)  not all  metals show 
a Drude-like behaviour and (ii)  such a behaviour 
can also be found in certain semiconductors \cite{AsM} and in new
artificial structures \cite{PHS}. 
Fcc structures of metallic spheres exhibit several exceptional 
properties \cite{AlM}.
For frequencies within $0.6 \omega_p \leq \omega\leq 1.1 \omega_p$,
where the bulk metal absorption can be negligible \cite{JC},
a CPBG opens in the spectrum with $g_w$ up to $\approx 10\%$,
and that already for a host dielectric constant 
$\varepsilon_h =1$. Moreover, up to {\em four} CPBG's can open 
in the spectrum.
A CPBG with $g_w\approx 5\%$ can be achieved for sphere 
filling fractions from $f=0.56$ till the close-packed case 
($f\approx 0.74$). 
For the wavelengths $\lambda$ within the CPBG's
the size parameter of spheres $x=2\pi r_s/\lambda$ satisfies $x\geq 5$.
Consequently, absorption is governed entirely by the bulk absorption,
since the so-called plasmon-induced absorption is negligible \cite{BH}.
Another unexpected feature is that for some values
of $r_s/\lambda_p$, for example $r_s/\lambda_p=1.35$,  
extremely narrow almost dispersionless bands `within a bandgap'
appear. These bands can to a certain extent perform many functions of 
an impurity band \cite{Y} since they involve photons with extremely small 
group and phase 
velocities (less than $c/200$) \cite{AlM}.

{\em Motivation}. - 
In this letter we study the question how the photonic-bandgap structure 
is affected by coating monodisperse metallic microspheres with a 
semiconductor or an insulator. This question is not only of 
theoretical but also of significant experimental interest. 
Coating can actually facilitate the preparation of photonic
colloidal crystals made up from metallic spheres because it can 
(i) stabilize metallic microparticles by preventing, or, at least, by 
significantly reducing their oxidation;
(ii) prevent aggregation of metallic particles by reducing Van der Waals
forces between them. In the latter case, a coating of roughly $30$ nm
is required. Also a suitable coating can enlarge some of 
the stop gaps (gaps in a fixed direction of propagation) by 
as much as $50\%$ \cite{MS}. 
From the application point of view, 
optically nonlinear Bragg diffracting nanosecond optical switches 
have been fabricated
by doping silica (SiO${}_2$) shell with an absorptive dye
\cite{PKA}. On the other hand, coating with an optically nonlinear 
material 
can reduce the required intensity for the onset of optical bistability 
\cite{KML} due to the enhancement of local 
fields near the surface-plasmon
resonance. 
Last but not least, using a semiconductor coating may allow a matching
of the photonic and electronic bandgaps, which is important for many
applications involving photonic crystals \cite{Y}.

Let $r_c$ be the core  radius and $l_c$ be the coating width, 
i.e., $r_s=r_c+l_c$. We assumed that the refractive indices 
$n_c$ and $n_h=\sqrt{\varepsilon_h}$ of the coating material and host, 
respectively, are  constant within the frequency range considered. 
The latter was taken to be roughly 
$0.55 \omega_p\leq  \omega\leq 1.1 \omega_p$, where the bulk 
absorption of the metal is assumed to be small.
This is a good approximation, for instance, for structures made of 
silica coated silver microspheres \cite{ULM}.
For silver $\lambda_p=328$ nm and the bulk absorption is 
rather small in  the region $310-520$ nm \cite{JC}. The dependence of the 
refractive index of silica
 on frequency in this frequency region is very weak and 
is described by a Cauchy model. The actual value of $n_c$ depends 
on the method used to synthetize silica. We took 
$n_c=1.47$. In view of their experimental relevance, we only 
investigated simple fcc structures.
Band-structure calculations were performed
using a photonic analogue \cite{Mo} of the familiar 
Korringa-Kohn-Rostocker  (KKR) method \cite{KKR}.
Compared to the plane-wave method,
dispersion does not bring any difficulties 
to the KKR method and computational time is the same as without
dispersion.
In order to ensure precision within $0.1\%$, spherical
waves were included with the angular momentum up to $l_{max}=10$.
Further discussion of convergence and errors can be found in
\cite{MS}. 
The values of the angular frequency are, unless otherwise 
stated, in the units $2c/A$,  
where $A$ is the length of the conventional unit cell 
of a cubic lattice (not to be confused with the lattice spacing \cite{AsM}).

\begin{figure}[tbp]
\begin{center}
\epsfig{file=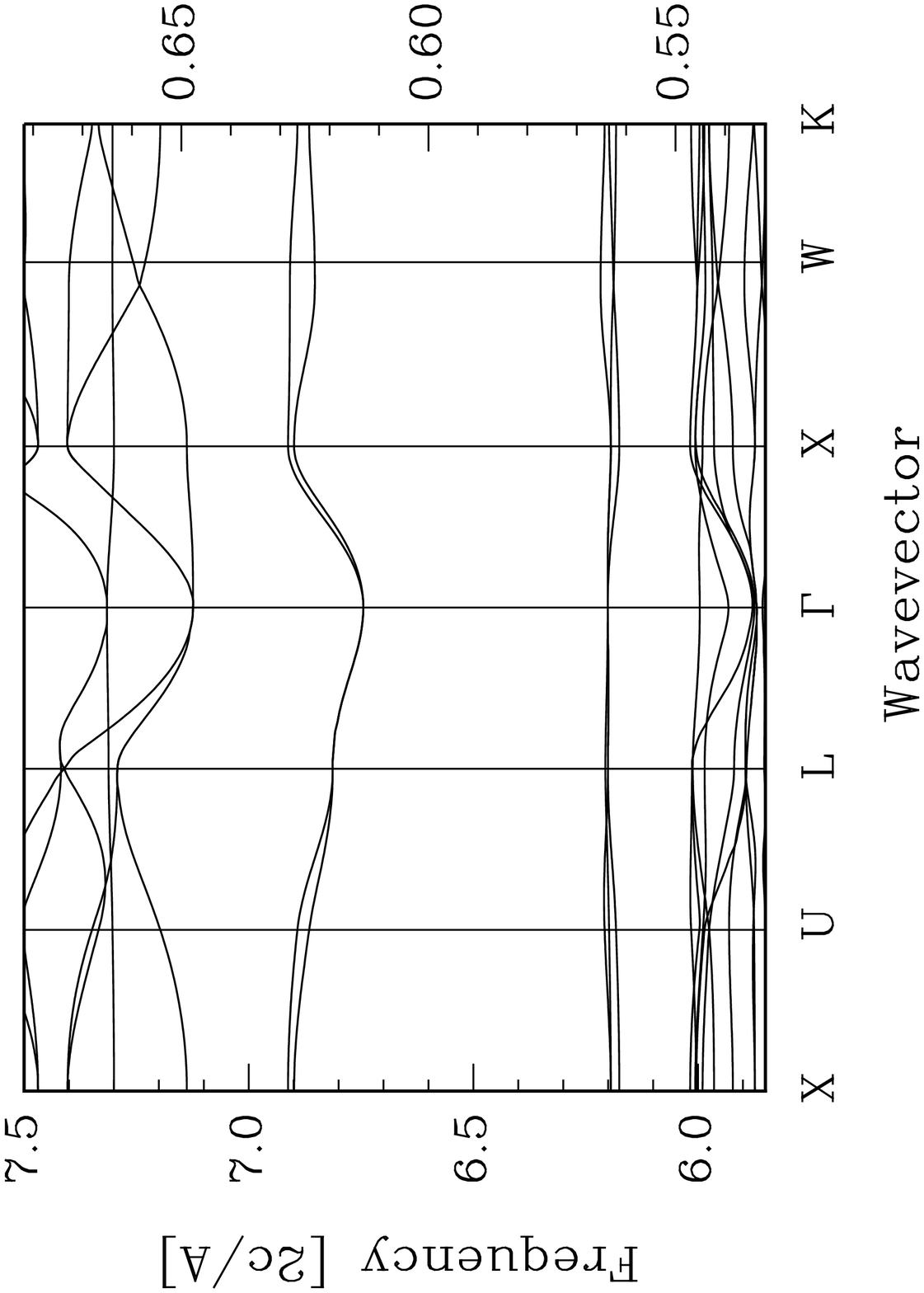,height=20cm,angle=-180}
\end{center}
\caption{}
\label{coated}
\end{figure}

\begin{figure}[tbp]
\begin{center}
\epsfig{file=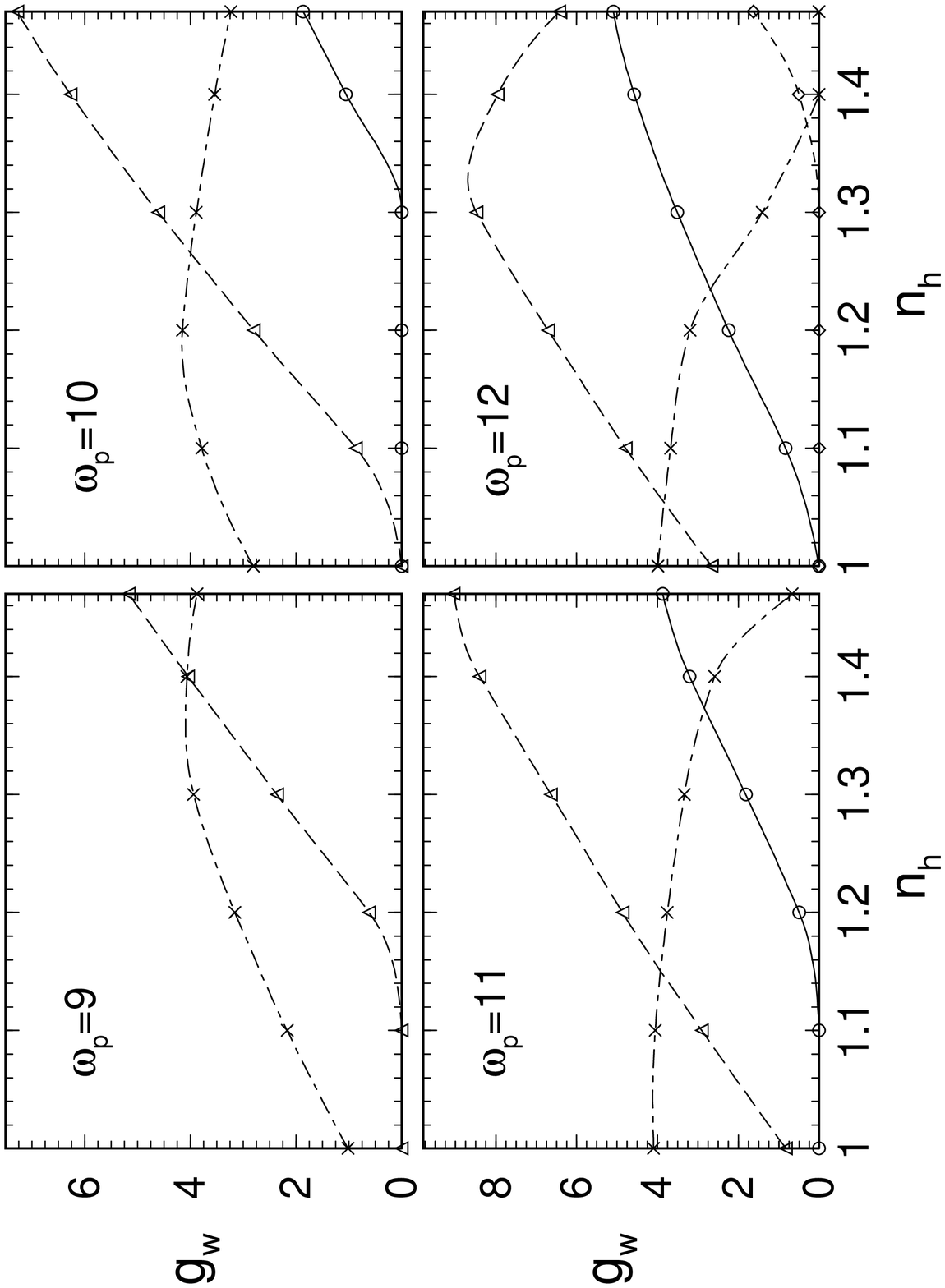}
\end{center}
\caption{}
\label{rgtot}
\end{figure}

{\em Results}. - 
At first glance it seems that coating destroys the CPBG's.
For example, if one begins with noncoated metallic spheres 
with $\omega_p=9.5$ and filling fraction $f=0.6$, 
one finds two CPBG's with $g_w\approx 4.07\%$ and $2.94\%$ at
midgap frequencies $\omega_c\approx 0.874 \omega_p$ and $0.813\omega_p$, 
respectively. As the coating width $l_c$ for $n_c=1.47$ increases 
from zero till the spheres are close-packed, the two CPBG's steadily 
decrease to zero. Nevertheless, when also the host
refractive index is allowed to vary, 
one can recover almost all the
exceptional features of the photonic band structure of 
noncoated metallic spheres. Generically, three CPBG's
open in the spectrum (see Fig. \ref{coated}), i.e., one less than
for noncoated metallic spheres, however, still two CPBG's more than
for purely dielectric structures \cite{HCS,BSS,MS}.
The relative gap width $g_w$ can exceed $9\%$.
Fig. \ref{rgtot} shows how the CPBG's of the close-packed
fcc lattice of silica coated metallic spheres 
with a coating width $l_c/\lambda_p  \approx  9.15 \%$ 
(corresponding to $l_c=30$ nm for silver), change when the host refractive
index $n_h$ is varied between $1$ and the coating refractive
index $n_c=1.47$.
Data on the right y-axis correspond to the case where $n_h$
matches $n_c$, i.e., they are identical to those for an fcc structure
of purely metallic spheres embedded in the host $n_h$.
It is clear from Fig. \ref{rgtot} that the answer to the question
whether coating enhances the relative gap width $g_w$ of a CPBG
strongly depends on $n_h$ and $r_s/\lambda_p$.

Let us denote $\omega_r$ as the midgap to plasma frequency
ratio, i.e., $\omega_r= \omega_c/\omega_p$.
Surprisingly enough, the ratio $\omega_r$ (for the CPBG's
shown in Fig. \ref{rgtot}) as a function of
$n_h$  can be described with 
high precision by a simple linear relation 
\begin{equation}
\omega_r(n_h)= \omega_r(n_0)-C(n_h-n_0),
\label{srel}
\end{equation}
with a universal constant $C=0.2478\pm 0.0075$. 
Here $n_0$ is the lowest
host refractive index for which a given CPBG opens in the spectrum
and $\omega_r(n_0)$ is the corresponding value of $\omega_r$
for $n_h=n_0$. This show that changing $n_h$ allows one
to tune not only the width $g_w$ of a CPBG but also, within
$\approx 8\%$, the corresponding midgap frequency.

\begin{figure}[tbp]
\begin{center}
\epsfig{file=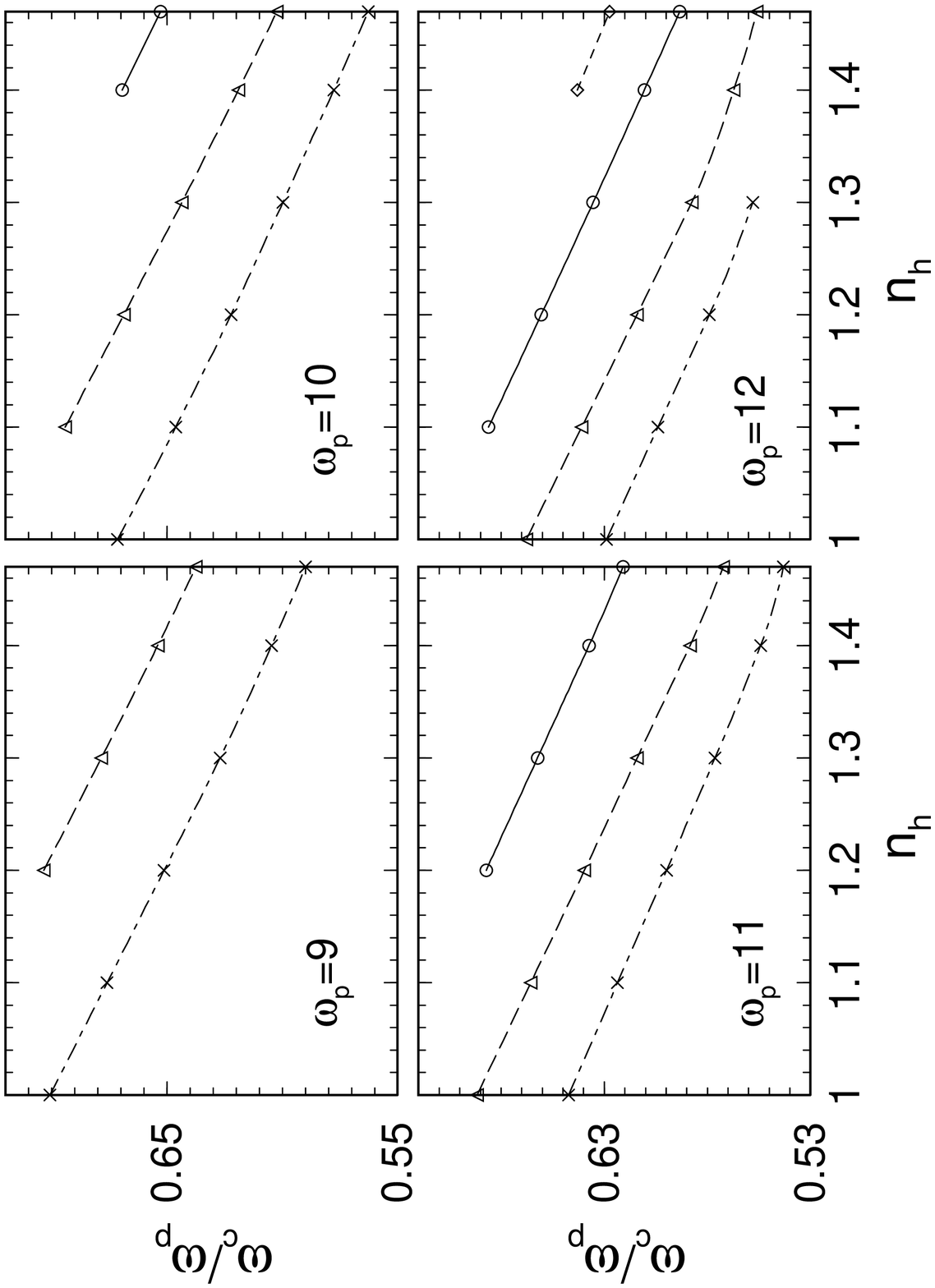}
\end{center}
\caption{}
\label{gfretot}
\end{figure}

The tuning of the midgap frequency $\omega_c$ by changing $n_h$
is more pronounced in the absence of coating.
In the latter case, for example for sphere filling fraction $f=0.6$,
$\omega_p=9$, and  $n_h=1$, two CPBG's appear with 
$g_w=3.27\%$ and $g_w=2.55\%$
at the midgap frequencies  $\omega_c\approx 0.904 \omega_p$ 
and $0.839\omega_p$, respectively. As $n_h$ increases from $1$ to $1.47$,
the midgap frequencies can be tuned down to $\approx 65\%$ of their
original values and their respective values  reach $\omega_c\approx 
0.593 \omega_p$ and $0.556\omega_p$. At the same time, the relative gap width
gradually increases up to $g_w=5.1\%$ and $g_w=6.43\%$, respectively
(see Figs. \ref{rgtot}, \ref{gfretot}).

The above results are not
specific for the case of silica coating when $n_c=1.47$. 
For example, for $n_c=1.4$ 
and $l_c/\lambda_p \approx 9.15\%$ one can find a region of 
parameters ($n_h=1.3$ and $\omega_p=12$ 
(the metallic filling fraction $f_m\approx 0.6$))
for which $g_w$ can be as large as $8.9\%$. 

It is important to realize that, because the metallic core size
parameter $x=2\pi r_c/\lambda$ satisfies $x\geq 3.4$ for
all wavelengths within a CPBG for all CPBG's considered here
in the frequency region $0.55 \omega_p\leq  \omega\leq 1.1 \omega_p$, 
the absorption is still 
dominated by bulk properties, i.e., can be negligible \cite{JC}, 
since the plasmon-induced absorption  becomes 
relevant only for particle sizes much smaller 
than the wavelength \cite{BH}.
One cannot get rid of absorption completely. Nevertheless
moderate absorption was shown to
cause only a slight perturbation of the band structure calculated  in 
the absence of absorption \cite{KMP1}.

It is worthwhile to mention that the  exact Drude-like dispersion 
(\ref{drudeps}) of $\varepsilon_s$ is not necessary to reproduce 
the exceptional properties of metallo-dielectric photonic crystals. 
It is enough if, for sufficiently large frequency window, 
 $-15  \stackrel{\textstyle <}{\sim} \varepsilon_s(\omega)\leq 0$  \cite{AlM}.
Many of the above features
(except for the extremely narrow almost dispersionless bands 
`within a bandgap') can also be reproduced for a constant and sufficiently
small negative $\varepsilon_s$ \cite{AlM}.
This is important  since, in real systems, a deviation
from the ideal Drude behaviour can occur at a proximity of the zero crossing of  
Re $\varepsilon$ at some $\lambda_z$. If $\lambda_p$
is  the plasma wavelength extracted from the fit (\ref{drudeps}) to 
a material data, then $\lambda_z$ is red-shifted compared to 
$\lambda_p$ \cite{JC} and the  band structure between $\lambda_z$ 
and $\lambda_p$ can be modified compared to the ideal Drude behaviour
(\ref{drudeps}).
 
{\em Outlook and conclusions}. -
Our calculations show that simple fcc structures of both metallic
and coated metallic spheres are ideal candidates to achieve
a CPBG for optical wavelengths. 
For coated microspheres with a coating width 
$l_c/\lambda_p \leq 10\%$
(up to $l_c=30$ nm for silver) and the coating and
host refractive indices $n_c$ and $n_h$, respectively,  
between $1$ and $1.47$, one can always find a
sphere radius $r_s$ such that the relative gap width $g_w$  is larger
than $5\%$ and, in some cases,  $g_w$ can even exceed $9\%$.
This provides a sufficiently large margin 
for gap-edge distortions due to omnipresent imperfections and impurities
to allow both technological and experimental
applications involving the proposed structures.  
Using different coatings and by changing the
refractive index $n_h$ of the supporting liquid 
(this can be easily achieved), one can tune the width and midgap frequency
of a CPBG considerably. In principle, the midgap frequency
$\omega_c$ can be tuned to whatever frequency within a 
nonabsorptive window ($0.6\omega_p\leq \omega \leq 1.1\omega_p$
for silver \cite{JC}).
Using a procedure  in which fluorescent 
organic groups are placed inside the silica shell with nm control over the 
radial position \cite{AvBV}
makes it, in principle, possible to perform  a precise 
position-dependent testing of the spontaneous emission within a 
photonic crystal. By applying an 
electric field one can switch in ms from an fcc colloidal crystal to a 
body centered tetragonal (bct) crystal: a so-called martensitic 
transition \cite{AvB}. Hence, the proposed structures are also
promising candidates for the CPBG structures
with tunable bandgaps. Last but not least, since metals are known to possess
large nonlinear susceptibilities, switching \cite{PKA}
and optical bistability \cite{KML} can be studied in the presence
of a CPBG. It is interesting to note that many of these ideas also applies
to two-dimensional photonic structures  \cite{LM}.

The region of plasma frequencies of conventional materials
ranges from the near-infrared to the ultraviolet \cite{BH}.
However, in a recent interesting paper \cite{PHS},
it has been shown that a whole new class of 
artificial materials  can be fabricated in which the plasma frequency may be reduced
by up to 6 orders of magnitude compared to conventional 
materials, down to GHz frequencies. Correspondingly, the proposed 
structures can provide CPBG structures from the GHz up to
ultraviolet frequencies.
Apparently, the main experimental problem in fabricating the 
proposed photonic  structures,  using colloidal systems of metallic
microspheres, is to synthetize large enough spheres
in order to reach the threshold value
$r_s n_h/\lambda_p \stackrel{\textstyle >}{\sim} 0.9$ to open a CPBG.
However, a method to produce monodisperse gold colloids 
of several hundred nm radius and larger has been developed \cite{GMa}.
Recent results on the fabrication of such spheres from silver, i.e.,
material with a  Drude-like behaviour of the dielectric function, 
are promising \cite{KPV}. The only remaining problem is to control 
the size polydispersity of spheres and reduce it below
$5\%$ to trigger crystalization \cite{RSS}.

I should like to thank  Dr. H. Bakker, Prof. A. van Blaaderen,  
Dr. A. Tip, and Dr. K. P. Velikov for careful reading of the manuscript 
and useful comments and M.J.A. de Dood for help with figures.
SARA computer facilities are also gratefully acknowledged.
This work is part of the research program by  the Stichting voor 
Fundamenteel Onderzoek der Materie  (Foundation for Fundamental 
Research on Matter) which  was made possible by financial support from the 
Nederlandse Organisatie voor Wetenschappelijk Onderzoek 
(Netherlands Organization for Scientific Research).

\newpage


\newpage

\begin{center}
{\large\bf Figure captions}
\end{center}

\vspace*{2cm}

\noindent {\bf Fig. 1.-} Calculated photonic-band-structure of a close-packed 
fcc lattice of silica coated ($n_c=1.47$) 
metallic spheres embedded in a host 
dielectric with refractive index $n_h=1.4$
for $\omega_p=11$ ($r_s/\lambda_p \approx 1.238$). 
For convenience, on the right y-axis the angular frequency is shown 
in units $\omega_p$.
The coating width $l_c$ to the sphere radius ratio is 
$l_c/r_s\approx 7.39\%$ ($l_c/\lambda_p\approx 9.15\%$).
This corresponds to a metallic (core) filling fraction 
of $f_m\approx 0.588$ (to be compared with the fcc close-packed
filling fraction $f\approx 0.7405$). 
Note that there are three CPBG's, one with $g_w=3.2\%$ at midgap 
frequency $7.01$, the second with $g_w=8.4\%$ at midgap frequency $6.47$,
and the third with $g_w=2.59\%$ at midgap frequency $6.1$.

\vspace{0.6cm}

\noindent {\bf Fig. 2.-} The relative gap width $g_w$ 
of the CPBG's of a close-packed fcc lattice of silica 
coated metallic spheres as a function of the host refractive index $n_h$
for $l_c/\lambda_p \approx 9.15\%$ (i.e., $l_c=30$ nm for silver). 
For a close-packed fcc lattice, the value of $r_s/\lambda_p$ 
can be recovered by multiplying the numerical values of $\omega_p$
by $1/(2\pi\sqrt{2})$, i.e., 
$r_s/\lambda_p=1.013$, $ 1.125$, $1.238$, and $1.35$ for 
$\omega_p=9$, $10$, $11$,  $12$ in this order.
The metallic filling fractions in the same order are
$f_m\approx 0.557$, $ 0.574$, $0.588$, and $0.6$.
The symbols  {\large$\circ$}, $\triangle$, and  {\large$\times$} are for the 
upper, middle and the lower CPBG, respectively (cf. Fig. \ref{gfretot}).
Data on the right y-axis correspond to the case when $n_h$
matches $n_c$.

\vspace{0.6cm}

\noindent {\bf Fig. 3.-}  The ratio $\omega_c/\omega_p$, corresponding 
to the CPBG's shown in Fig. \ref{rgtot}, as a function of 
the host refractive index $n_h$.
With a high precision the curves can be described
by a simple linear relation [see Eq. (\ref{srel})].

\newpage

\begin{thebibliography}{99}

\bibitem{OG}Graydon O., OLE June 1998, p. 31.

\bibitem{Y}Yablonovitch E.,
Phys. Rev. Lett. {\bf 58} (1987) 2059.

\bibitem{HCS}Ho K. M., Chan C. T. and Soukoulis C. M.,
Phys. Rev. Lett. {\bf 65} (1990) 3152.

\bibitem{YGL}Yablonovitch E., Gmitter T. J. and Leung K. M.,
Phys. Rev. Lett. {\bf 67} (1991) 2295.

\bibitem{AMe}Moroz A., Europhys. Lett. {\bf 46} (1999) 419.

\bibitem{Tip}Glauber R. J. and Lewenstein M.,  
Phys. Rev. A {\bf 43} (1991) 467,
Eq. (6.12); Tip A., Phys. Rev. A {\bf 56} (1997) 5022;
ibid. {\bf 57} (1998) 4818.

\bibitem{WW}Weisskopf V. and Wigner E., Z. Phys. {\bf 63} (1930) 54.

\bibitem{By}Bykov V. P., Sov. J. Quant. Electron. {\bf 4} (1975) 861.

\bibitem{LZM}Lewenstein M., Zakrzewski J., Mossberg T. W. and Mostowski J.,
J. Phys. B{\bf 21} (1988) L9.

\bibitem{RSS}Russel W. B., Saville D. A. and Schowalter W. R.,
{\em Colloidal Dispersions} (Cambridge University Press, Cambridge, 1995).

\bibitem{WV}Wijnhoven J. E. J. G. and Vos W. L.,
Science {\bf 281} (1998) 802.

\bibitem{BSS}Biswas R., Sigalas M. M., Subramania G. and Ho K.-M., 
Phys. Rev. B {\bf 57} (1998) 3701.

\bibitem{MS}Moroz A. and Sommers C., 
J. Phys.: Condens. Matter {\bf 11} (1999) 997. 

\bibitem{Lev}Levi B. G., Phys. Today, January 1999, p. 17.

\bibitem{Lin}Lin S. Y. et al., Nature {\bf 394} (1998)  251.

\bibitem{AlM}Moroz A., Phys. Rev. Lett. {\bf 83} (1999) 5274.

\bibitem{BH}Bohren C. F. and Huffman D. R., {\em Absorption
and Scattering of Light by Small Particles}
(Wiley, New York, 1984), Chap. 9, 12.

\bibitem{AsM}Ashcroft N. W. and Mermin N. D., {\em Solid State Physics}
(Saunders College, 1976) pp. 73-75.

\bibitem{PHS}Pendry J. B., Holden A. J., Stewart W. J. and Youngs I.,
Phys. Rev. Lett. {\bf 76} (1996) 4773.

\bibitem{JC}Johnson P. B. and Christy R. W.,
Phys. Rev. B {\bf 6} (1972) 4370; ibid. {\bf 9} (1974) 5056.

\bibitem{PKA}Pan G., Kesavamoorthy R. and Asher S. A.,
Phys. Rev. Lett. {\bf 78} (1997) 3860;
Grier D., Phys. World, July 1997, p. 24.

\bibitem{KML}Leung K. M., Phys. Rev. A {\bf 33} (1986) 2461;
Chemla D. S. and Miller D. A., Opt. Lett. {\bf 11} (1986) 522.

\bibitem{ULM}Ung T., Liz-Marzan L. M. and Mulvaney P., 
Langmuir {\bf 14} (1998) 3740.

\bibitem{Mo}Moroz A., 
Phys. Rev. B {\bf 51} (1995) 2068.

\bibitem{KKR}Korringa J.,
Physica {\bf 13} (1947) 392; Kohn W. and Rostoker N.,
Phys. Rev. {\bf 94} (1954) 1111. 

\bibitem{KMP1}Kuzmiak V. and Maradudin  A. A.,
Phys. Rev. B {\bf 55} (1997) 7427.

\bibitem{AvBV}van Blaaderen A. and Vrij A., 
Langmuir {\bf 8} (1992)  2921.

\bibitem{AvB}van Blaaderen A., MRS Bulletin {\bf 23}, October (1998) 39;
Dassanayake U., Fraden S., and  van Blaaderen A.,
J. Chem. Phys. {\bf 112} (2000) 3851.

\bibitem{LM}van der Lem H. and Moroz A., to appear in J. Opt. A: Pure Appl.

\bibitem{GMa}Goia D. V. and Matijecic E., 
New J. Chem. {\bf 146} (1999) 139.

\bibitem{KPV}Velikov K. P., private communication.


\end{thebibliography}
\end{document}